# Intercalation of metal into transition metal dichalcogenides in molten salts

Lin Gao,[1,2,3] Mian Li,[1,3,*] Binjie Hu[2,**], and Qing Huang [1,3,***]

[1]Engineering Laboratory of Advanced Energy Materials, Ningbo Institute of Materials Technology and Engineering, Chinese Academy of Sciences, China

[2]Department of Chemical and Environmental Engineering, Faculty of Science and Engineering, University of Nottingham, China

[3]Qianwan Institute of CNiTECH, China

*Correspondence: limian@nimte.ac.cn, binjie.hu@nottingham.edu.cn, huangqing@nimte.ac.cn

**Abstract:** Van der Waals (*vdW*) layered materials have drawn tremendous interests due to their unique properties. Atom intercalation in the *vdW* gap of layered materials can tune their electronic structure and generate unexpected properties. Here we report a chemical-scissor mediated method that enables metal intercalation into transition metal dichalcogenides (TMDCs) in molten salts. By using this approach, various guest metal atoms (Mn, Fe, Co, Ni, Cu, and Ag) were intercalated into various TMDCs hosts (such as $TiS_2$, $NbS_2$, $TaS_2$, $TiSe_2$, $NbSe_2$, $TaSe_2$ and $Ti_{0.5}V_{0.5}S_2$). The structure of the intercalated compound and intercalation mechanism was investigated. The results indicate that the *vdW* gap and valence state of TMDCs can be modified through metal intercalation, and the intercalation behavior is dictated by the electron work function. Such a chemical-scissor mediated intercalation provides an approach to tune the physical and chemical properties of TMDCs, which may open an avenue in functional application ranging from energy conversion to electronics.

# Introduction

Two-dimensional (2D) materials have attracted significant attention owing to their unparalleled properties, such as thermoelectric and optoelectronics [1-3]. Weak van der Waals (*vdW*) force in the 2D materials endows the possibility of intercalation of guest species into gaps between sublayers [4,5]. Such intercalation offers a flexible approach for tuning the properties of 2D materials by formation of longitudinal heterogeneous structures, leading to potential applications in energy storage, electronics, photonics, etc. [6-12]. Transition metal dichalcogenides (TMDCs) with common formula of $MX_2$ (M is a transition metal of groups 4–10; X is S, Se or Te) where a layer of M is sandwiched between two layers of chalcogens atoms, are typical 2D *vdW* materials. The intercalation of TMDCs has been proven an effective approach to modify their physical and chemical properties[13].

Intercalation usually results in change of valence state of elements in 2D materials [14], in which the Fermi level of hosts moved up due to filling of extra electrons in the empty *d* orbital [15]. In general, the well-accepted intercalation approach involves vapor transport technique [16,17], wet-chemical methods [18], electrochemical intercalation [19-21], etc. However, some intercalation compounds that are not thermodynamically stable phases, are hard to be obtained through these approaches. Therefore, alternative methods should be developed to further explore unprecedented properties of 2D materials.

Recently, we revealed that the interlayer atoms in nanolaminated carbides and nitrides, the so-called MAX phases, can be substituted by either metals or anions in Lewis acidic molten salts [22,23]. This method could provide more terminal options for 2D MXenes than the HF etching method, such as O, Se and Te through anion-exchange reaction in melts [22,23]. Moreover, melted metals can served as chemical scissors to remove the terminals on the surface of MXenes and accomplished the reconstruction of MAX phases [24]. Therefore, chemical scissor-mediated structural editing protocol has great potential to accurately tailor 2D materials such as gap-engineering and intercalation of guest species.

Herein, we use chemical scissor-mediated structural editing protocol to intercalate metal atoms into 2D TMDCs in molten salts. In contrast to the electron acceptor nature of Lewis acid molten salts accounting for the etching of MAX phases, reductive metals in molten salts can act as electron donors, which promote opening the *vdW* gap of TMDCs and the subsequent metal intercalation. Various guest atoms (Mn, Fe, Co, Ni, Cu, and Ag) were intercalated into a series of TMDCs (such as $TiS_2$, $NbS_2$, $TaS_2$, $TiSe_2$, $NbSe_2$, $TaSe_2$ and $Ti_{0.5}V_{0.5}S_2$) in LiCl-KCl molten salt. Density Functional Theory (DFT) was also performed to illustrate the intercalation mechanism.

## Results and discussion

### Evidence of intercalation: lattice expansion and superspots

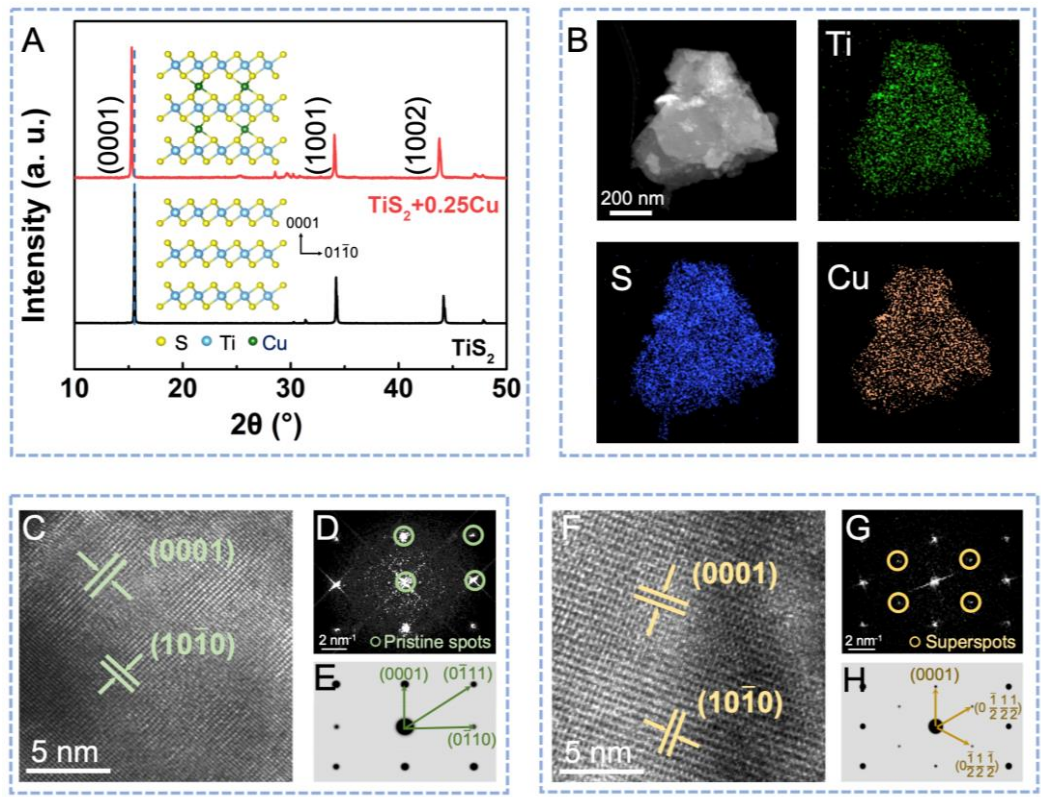

Figure 1. Structural characterization of $TiS_2$ and Cu intercalated $TiS_2$

(A) XRD patterns of pristine $TiS_2$, and Cu intercalated $TiS_2$, insets show the atomic structure of $TiS_2$ and $Cu_{0.25}TiS_2$. (B) HADDF-EDS elemental mappings of Ti, S and Cu in $Cu_{0.25}TiS_2$. (C) HRTEM images of pristine $TiS_2$ and (D) their corresponding selected area diffraction along the $[01\bar{1}0]$ zone axis as well as (E) simulated pattern. (F) HRTEM images of pristine $Cu_{0.25}TiS_2$, and (G) their corresponding selected area diffraction along the $[01\bar{1}0]$ zone axis as well as (H) simulated pattern.

The structure features of original TiS$_2$ and Cu-intercalated TiS$_2$ *via* chemical scissor-mediated structural editing protocol were characterized with various techniques as Figure 1 shows. **Figure 1A** exhibits XRD (000*l*) peaks of Cu-intercalated TiS$_2$ shifting towards lower angles comparing to that of TiS$_2$, indicating that the expanded *c*-axis lattice after Cu intercalation [25]. In **Figure 1B**, Cu element is uniformly distributed in the TiS$_2$ lamellae, and HAADF-EDS semi-quantitative analysis (**Figure S1**) shows the molar ratio of Cu: Ti: S is 8.1: 32.3: 59.6 (at %), indicating the molecular formula of Cu$_{0.25}$TiS$_2$. The HRTEM image in **Figure S2** reveals the local structure changes after intercalation along the [0001] zone-axis direction. Note that these stripes in **Figure S2B** are superstructure instead of Moiré fringes (because Moiré fringes caused by the overlap of pristine or intercalated phases, should have lattice stripes approximately 20 times larger), which are frequently observed in intercalation [26]. Fast Fourier Transform (FFT) pattern for pristine TiS$_2$ in **Figure 1D** indicates the viewing direction in **Figure 1C** is [01$\bar{1}$0]. The corresponding simulated electron diffraction patterns of TiS$_2$ is also shown in **Figure 1E**. While in **Figure 1F**, the lattice fringes spacing along *c* axis were slightly enlarged due to intercalation. The angle between the (10$\bar{1}$0) and (0001) crystal planes is still 90°, which means the pristine structure remains. The FFT pattern in **Figure 1G** shows new diffraction superspots (marked with yellow circle), which suggests the long-range ordering of intercalated Cu atoms. This periodicity in electron diffraction is due to the fact that the intercalated Cu atoms minimizes the geometrical mismatch between the two lattices by periodic deformation to obtain optimum energy interaction [27]. This structural evolution can be described by the following types of structures: a $\sqrt{3a} \times \sqrt{3a}$ superstructure of Cu$_{0.25}$TiS$_2$, which is composed of two different periodicities of TiS$_2$ and Cu. Moreover, the newly crystalline plane angle with the (0$\bar{1}$10) plane of TiS$_2$ is about 28.6°, and its corresponding spacing is 0.602 nm.

## Mechanism of intercalation: chemical scissor in molten salts

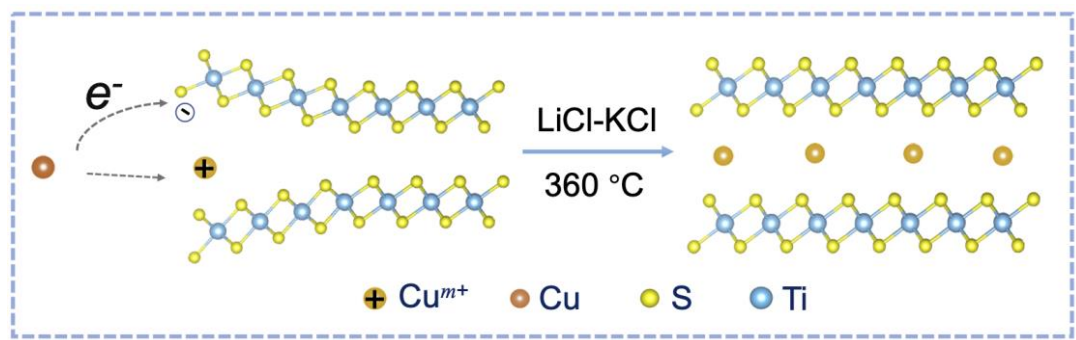

Figure 2. Intercalation schematic of Cu-TiS$_2$ in LiCl-KCl eutectic molten salt at 360 °C

**Figure 2** shows the sketch of intercalation and the formation of Cu-TiS$_2$. Electrons (e$^-$) could be transiently loose in reductive metals and enter into melts [28]. These solvated electrons ($m$e$^-$) tend to form F-center structure surrounded by cations (Li$^+$ or K$^+$), which warranty a finite lifetime due to their strong coupling [29]. Meanwhile, the positively charged metals cations (Cu$^{m+}$) were desolvated into the molten salts driven by the polarization of Cl$^-$. As the solvated electrons were captured by acceptors in melts, the cations could interact with the negatively charged species *via* electrostatic force. Taking Cu-TiS$_2$ as an example, following reactions (Equations (1)-(3)) could occur during intercalation of Cu into TiS$_2$:

Cu = Cu$^{m+}$ + $m$ e$^-$                                                                  (1)

$m$ e$^-$ + TiS$_2$ = (TiS$_2$)$^{m-}$                                          (2)

(TiS$_2$)$^{m-}$ + Cu$^{m+}$ = CuTi$^{(4-m)}$S$_2$                          (3)

Solvated electrons have been confirmed to participate reduction reaction in solvents [30,31]. Therefore, as illustrated by reaction (1)-(2), these solvated electrons could reduce the Ti into TiS$_2$ host, and then negatively charge these layered materials. Such electron injection should open the *vdW* gap that mimics the removal of terminals of MXenes by metal scissors [30,31]. The positively charged Cu$^{m+}$ cations then diffuse into *vdW* gap driven by electrostatic force, to form Cu-intercalated TiS$_2$ uniformly. This speculation is supported by **Figure S3**, which shows the intermediate state of Cu-TiS$_2$ intercalation after 1 hour reaction. The intensity of HADDF shown in **Figure S3F** quantitatively confirms the lower concentration of Cu in the center of the TiS$_2$ particle, and the corresponding EDS analysis are present in **Figure S3**. Obviously that Cu enriches around the rim of the particle (49.9 at%) but is scarce in the core region (5.2 at%) in the early stage of intercalation. For Cu$_{0.25}$TiS$_2$, the consistency between the First-principles density-functional theory (DFT) calculation (d$_{0001}$=0.579 nm) and experimental results (d$_{0001}$=0.575 nm) suggest that Cu atoms are preferentially in the octahedral coordination site in *vdW* gaps of TiS$_2$ (see the inset of **Figure 1A**), which also agrees with previous study [32].

**Interaction of guest and host: ordering and electron transfer**

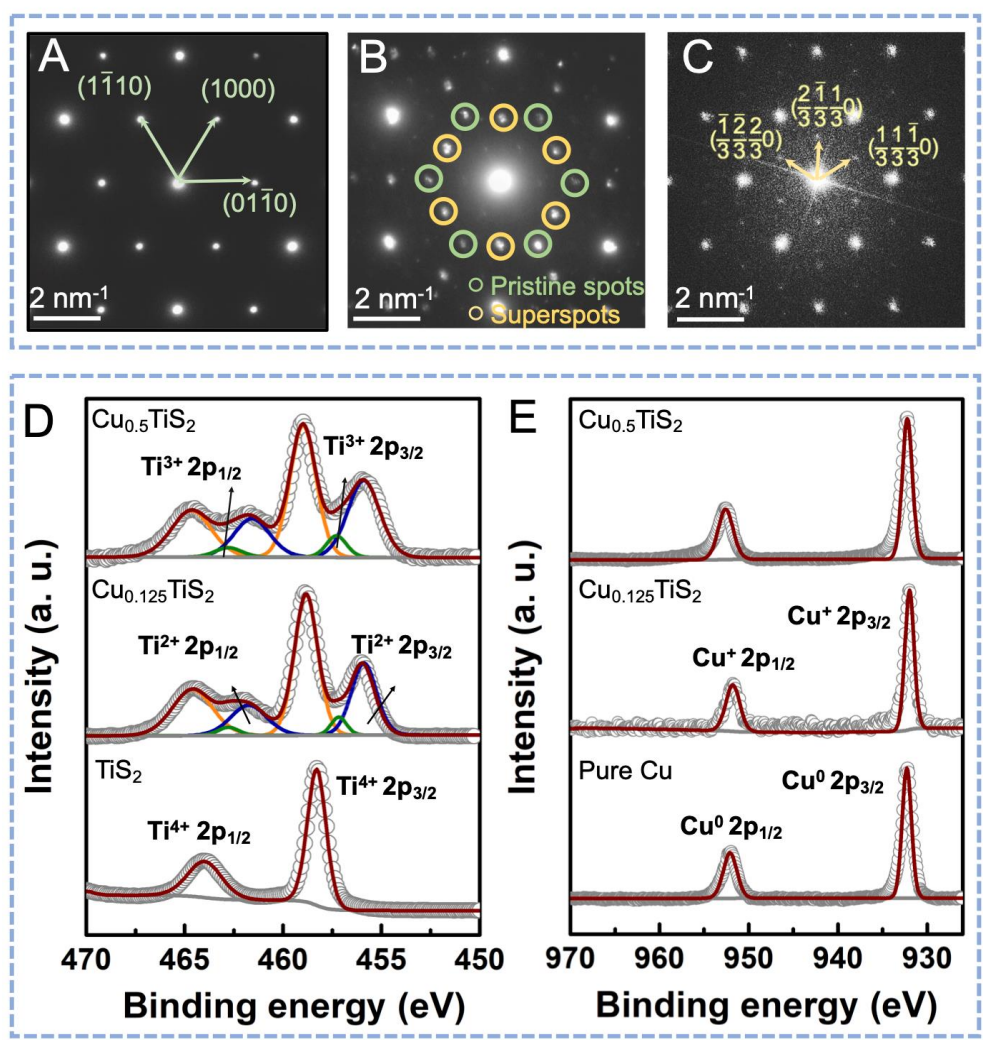

Figure 3. Superspots and charge transfer for various Cu intercalation concentrations (A-C) FFT patterns of pristine TiS$_2$ diffraction spots, Cu$_{0.125}$TiS$_2$ and Cu$_{0.5}$TiS$_2$ diffraction superspots taken along the [0001] zone-axis direction. (D-E) Ti 2p and Cu 2p XPS analysis of Cu$_x$TiS$_2$. ($x$= 0, 0.125, 0.5).

**Figure 3** show structure evolution and electron transfer behavior of Cu-intercalated $Cu_xTiS_2$ (x=0, 0.125, 0.5). **Figure 3A-C** demonstrate diffraction spots viewed along [0001] direction. The intermediate state superspots with a hexagon of 12-spots (newly: 6 midways between adjacent bright spots) are observed in $Cu_{0.125}TiS_2$ (**Figure 3B**). Associated periodic structural distortion occurs to minimize the repulsive interactions of guest Cu ions and reduce the strain energy of elastic coherence, indicating an incommensurate charge density wave signature [33]. While the final superspots in the $Cu_{0.5}TiS_2$ (**Figure 3C**) are distinctly different, where 6 newly spots are allocated in the centers of equilateral triangles of bright spots, inside the pristine hexagonal 6-spots. Such new periodicity also corresponding to a $\sqrt{3}$ a× $\sqrt{3}$ a superstructure accompanied with a 29.8° rotation, which is consistent with the superstructure observed along [01$\bar{1}$0] axis in $Cu_{0.25}TiS_2$ (**Figure 1G**). Overall, the transition in diffraction spots suggests that the intercalated Cu atoms in the *vdW* gap leads to various degree of ordering.

Moreover, electron transfer with various Cu intercalated are further investigated by X-ray photoelectron spectrum (XPS) analysis. Ti 2p and Cu 2p X-ray photoelectron spectrum of the as-synthesized samples shown in **Figure 3D-E**, indicate the bonding state differences of $Cu_xTiS_2$ after intercalation. As mentioned above, intercalation process should involve charge transfer between hosts and guest species. Transition metals with a partially filled *d*-band may accept electrons donated by the guest [34,35]. Thus, the concentration of Cu was controlled to investigate charge transfer behavior (XRD patterns and EDS analysis were presented in **Figure S4**). As shown in **Figure 3D**, only a pair of splitting peaks was found with Ti. This may attribute to Ti (IV) ($2p_{3/2}$) and Ti (IV) ($2p_{1/2}$) before intercalation. While for $Cu_xTiS_2$, the orbital state of Ti clearly shifts towards the lower valence, which means the external electrons have partially reduced the binding of Ti-S. Multiple valance peaks for Ti at 455.9 eV, 457.2 eV and 458.8 eV are assigned to the Ti-S (II) ($2p_{3/2}$), Ti-S (III) ($2p_{3/2}$) and Ti-S (IV) ($2p_{3/2}$) bonds. In addition, Cu peaks at 931.9 eV and 951.7 eV (**Figure 3E**), correspond to univalent Cu (I) doublet peaks as $2p_{2/3}$ and $2p_{1/2}$ orbital, respectively. Notably, no significant valance change was observed for Cu with increasing Cu intercalation concentration. This phenomenon is consistent with previous research, which confirms the weak charge transfer between Cu and TiS$_2$ during intercalation [36].

**Electron modulation of guest and host in intercalation**

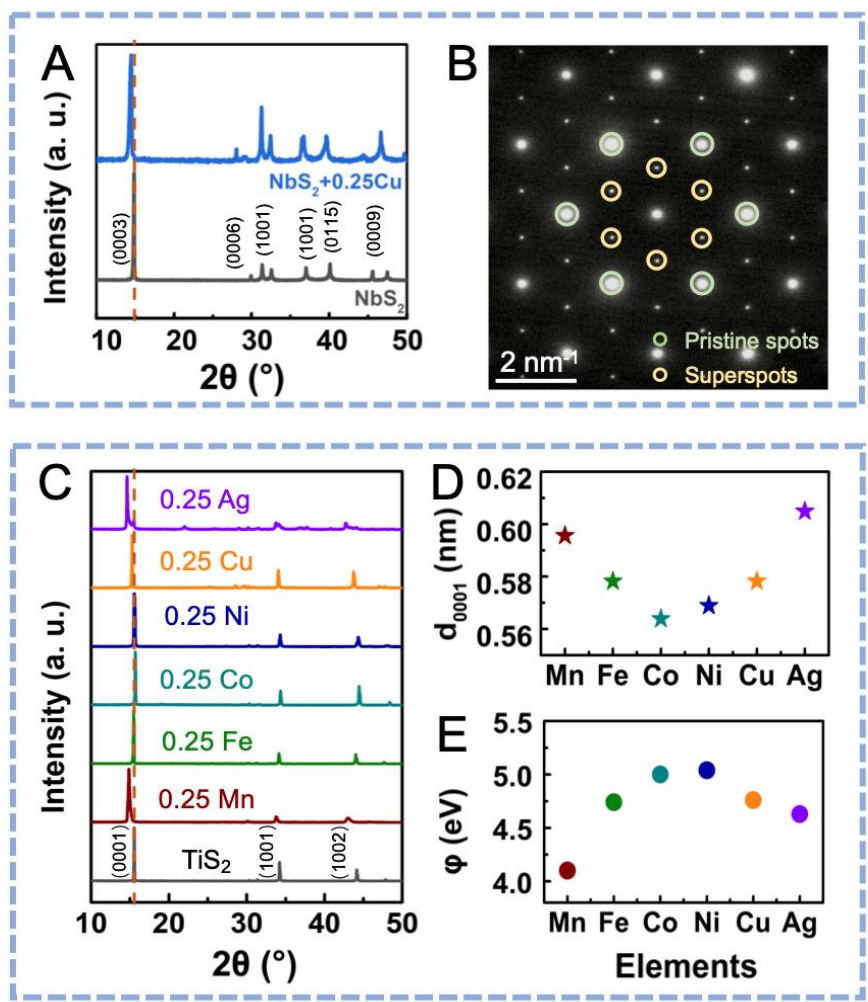

Figure 4. Electron modulation of guest and host in intercalation part 1 (A) XRD patterns of Cu$_{0.25}$NbS$_2$ and NbS$_2$. (B) SAED pattern of Mn$_{0.25}$TiS$_2$ diffraction superspots taken along the [0001] zone-axis direction, green circle: pristine spots; orange circle: superspots. (C) XRD pattern of A$_{0.25}$TiS$_2$ (A= Mn, Fe, Co, Ni, Cu, and Ag) obtained in solvated electron molten salt. (D,E) Correlation between interlayer spacing of I$_{0.25}$TiS$_2$ and electron work functions ($\varphi$, eV) of guest.

Similar behavior was detected in other intercalated compounds. The shifted XRD peaks of Cu-intercalated NbS$_2$ in **Figure 4A** demonstrates a similarly enlarged unit cell.

Intercalation in other hosts, such as TaS$_2$, TiSe$_2$, NbSe$_2$ and TaSe$_2$, can be found in **Figure S5-9**, which further illustrate the versatility of chemical scissor-mediated intercalation in molten salts. Moreover, most of TMDCs exhibit the maximum intercalation molar ratio of 0.5 in the final intercalation compounds Cu$_{0.5}$MX$_2$, except for host TaSe$_2$ which can only accommodate Cu up to Cu$_{0.20}$TaSe$_2$, indicating the partial-filling *d* orbital in TMDCs has important influence on the amount of intercalated guest species.

Additionally, different intercalation elements were investigated, including Mn, Fe, Co, Ni, Cu, and Ag. As shown in **Figure 4B**, the electron diffraction spots of Mn$_{0.25}$TiS$_2$ are comparable to the superspots observed in Cu$_{0.5}$TiS$_2$ (6 new spots are in the centers of equilateral triangles of bright spots). Furthermore, as shown in **Figure 4C**, the intercalation behavior of A$_{0.25}$TiS$_2$ (A= Mn, Fe, Co, Ni, Cu, and Ag, see **Figure S10-15** for details) in terms of layer spacing with previous report *via* vapor transport technique [16]. In previous studies, the intercalation has been analyzed in terms of ionic radius, bond length, and bond ionicity, etc. [16,37]. Given the shrinkage of the layer spacing cannot be fully explained by the ionic radius, also the trend of the ionic radii (Mn$^{2+}$: 0.083 nm; Fe$^{2+}$: 0.061 nm; Co$^{2+}$: 0.65 nm; Ni$^{2+}$: 0.69 nm; Cu$^+$: 0.77 nm; Ag$^+$: 1.15 nm) and the layer spacing do not exactly match with each other, leading to the introduction of bond length and bond ionicity. Here, we can analyse it in terms of electron redistribution at the interface of electron donor and acceptor. Actually, such ordered intercalation in the *vdW* gap results in the molar ratio of total positive and negative ions deviating from the original stoichiometric ratio, which disrupts the original electron distribution within the lattice. This fact allows excess electrons and positive charges (hole) emerging in local regions of the lattice. Such redistribution of the excess depends on electrons not only the electron bound ability of the acceptor, but also the electron supply ability of the donor. Generally, electron work function ($\varphi$) is the minimum energy required to remove an electron from the interior of a solid to a point just outside the surface, which affects the nuclei-electron and electron–electron interactions [38]. Although EWF is a surface parameter, it is related closely to the atomic electronic state, which can be used to evaluate their electron donating ability [39].

The relationship between binding states in final intercalation compounds and electronic state of intercalated atoms was evaluated. Results presented in **Figure 4D** and **4E**, indicates that the interlayer spacing of $A_{0.25}TiS_2$ is inversely related with the $\varphi$ of guest elements. The larger $\varphi$ obtained with Co and Ni suggest the strong electrostatic attraction result in the shrinkage of interlayer spacing, hence more electrons overlap in their bonding, which tends to be covalent in their bound state with S. The smaller $\varphi$ [Mn] leads to larger spacing of the intercalation compound due to the weak ionic bonding. Overall, the electron work function of the guest atom implies variable electron donating abilities in the electron transfer caused by different guest intercalation. Such electronic matching balance between guest and host determines whether the bound electrons are direct gain-loss of electrons (lead to ion-like bonding) or partial overlap of shared electrons in an extended electron cloud (lead to covalent-like bonding).

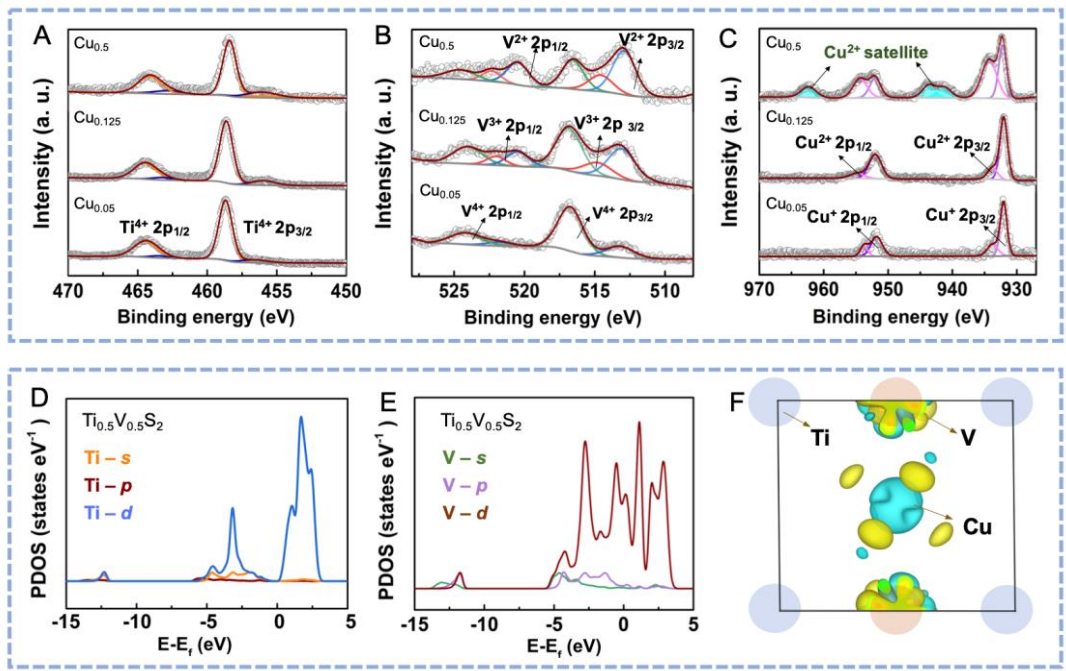

Figure 5. Electron modulation of guest and host in intercalation part 2 (A-C) Ti 2p, V 2p and Cu 2p XPS analysis of $Cu_xTi_{0.5}V_{0.5}S_2$, ($x$= 0.05, 0.125, 0.5); (D-E) Projected density of states of $Ti_{0.5}V_{0.5}S_2$ with Ti 3s, 3p, 3d and V 3s, 3p, 3d (details about $TiS_2$ see **Figure S18**). (F) Side views of charge density difference of $Cu_{0.25}Ti_{0.5}V_{0.5}S_2$ with [01$\bar{1}$0] viewing direction, yellow region: charge accumulation; blue region: charge depletion (details about $Cu_{0.25}TiS_2$ see **Figure S18**).

Furthermore, we believe that the electron accepting ability of the host (Ti is the acceptor for TiS$_2$) is also critical to the intercalation, thus we have regulated the electron acceptor by means of solid solution at M-site. V with different valence electrons was mixed in TMDCs to identify how the *d* electron density affects intercalation. **Figure 5** exhibits both experimental and theoretical calculation results of charge transfer when Cu was intercalated in Ti$_{0.5}$V$_{0.5}$S$_2$. Ti, V and Cu were found being evenly distributed in the final product after intercalation accompanied with lattice expansion (Ti$_{0.5}$V$_{0.5}$S$_2$ d$_{0001}$ =0.561 nm; Cu$_{0.25}$Ti$_{0.5}$V$_{0.5}$S$_2$ d$_{0001}$ =0.569 nm), the corresponding EDS analysis and XRD pattern is shown in **Figure S16**. **Figure 5A-C** show Ti 2p, V 2p and Cu 2p X-ray photoelectron spectrum of Cu$_x$Ti$_{0.5}$V$_{0.5}$S$_2$ (*x*=0.05, 0.125, 0.5). Notably, the electron acceptor was fundamentally changed, and the amount of electron being transferred was enhanced when mixing with V. For Cu$_x$Ti$_{0.5}$V$_{0.5}$S$_2$, the valence state of Ti is nearly unchanged compared to Cu$_x$TiS$_2$ (**Figure 3D**), indicating that a competition for the preferred electrons occurs and Ti is no longer the electron acceptor. While the valance of V has gradually shifted to lower state with emerged V (III) shoulder peak at 514.6 eV [40], and the proportion of low-valent V (II, III) expands with increasing Cu. This fact indicates that the partial-fill *d* orbital of V is preferred as the primary electron acceptor. As for Cu placed in the *vdW* gap, a strong satellite peak of Cu (II) has been motivated between 935-945 eV in **Figure 5C**, which indicates the enhanced electron donating and strengthen electron transfer. Overall, the previous acceptor and donor pair have been totally altered due to the electron acceptance competition of two host elements.

Additionally, to elucidate the altered electron redistribution at the interface between the guest and host, total density of state (TDOS) and the project density of state (PDOS) of the TiS$_2$ and Ti$_{0.5}$V$_{0.5}$S$_2$ are calculated. Compared with the PDOS of Ti-TiS$_2$ shown in **Figure S17**, all peaks of Ti-Ti$_{0.5}$V$_{0.5}$S$_2$ (**Figure 5D**) moved up to higher energy level, indicating an increasing orbital energy of the Ti. While the TDOS diagram of V (**Figure 5E**) shows broad peak around the Fermi level, implying a wide electron energy distribution for V. The delocalized electron distribution suggests a greater orbital overlapping for V-S bonds. Therefore, lower orbital energy is the reason why V is preferred by donated electrons as the primary acceptor in Cu-Ti$_{0.5}$V$_{0.5}$S$_2$. Besides, semi-quantitative electron transfer is illustrated by charge density differences (CDD) and Bader analysis shown in **Figure 5F**. Although a net gain of electrons is depicted in both cases for TMDCs, Ti can accept electron to varying degrees in Cu$_{0.25}$TiS$_2$ (**Figure S18**); whereas in Cu$_{0.25}$Ti$_{0.5}$V$_{0.5}$S$_2$, only V becomes the primary electron acceptor. These calculations are broadly accordant with the experimental results. Similarly, the Bader analysis results indicate that more charge (-0.61 eV) being lost with Cu than that of Ti (-0.57 eV) at M-site. It is clear that the electron transfer strength determines the electron accepting ability of host.

## Conclusion and outlook

The chemical scissor-medicated intercalation of TMDCs were realized in molten salts, which offers a versatile protocol to edit the structure and composition of *vdW* layered materials. Superlattices have been observed in different system and enables the valance state regulation of both guest and TMDCs. Moreover, molten salts with solvated electrons can serve as a bridge to connect adducts to atomic layer with empty orbitals, transforming the surface characteristics of the 2D materials without altering their lattice structure and topology. We foresee the potential impact of this approach on establishing unexpected physical behaviors.

## Experimental procedures

### Resource availability

Further information and requests for resources and materials should be directed to and will be fulfilled by the lead contact, Prof. Qing Huang (huangqing@nimte.ac.cn).

All materials generated in this study are available from the lead contact without restriction.

All data can be seen in the main text or supplemental information, including experimental and computational methods. Characterization data of all compounds. Including high-resolution TEM images, SEM image, EDS analysis, XPS data, and XRD data.

**Supplemental information**

Document S1. Supplemental experimental procedures, Figures S1–S18, and Table S1

Table S1. The composition and condition of starting materials for synthesizing $A_xMX_2$

Figure S1: Corresponding EDS analysis of Cu intercalated $TiS_2$ in Figure 1b

Figure S2: HRTEM image of (a) $TiS_2$ and (b) $Cu_{0.25}TiS_2$ taken along the [0001] zone-axis direction

Figure S3: (a-e) Elemental mapping of Cu intercalated $TiS_2$ at 1 h intermediate state; (f) line analysis of Ti and Cu at 1 h intermediate state; (g) dark-field microscopy of Cu intercalated $TiS_2$ at 1 h intermediate state, (h-i) EDS analysis at the edge area 1 and the center area 2

Figure S4: (a) XRD pattern of pristine $TiS_2$ and $Cu_xTiS_2$ ($x$=0.05, 0.125, 0.25, 0.5); (b-e) SEM image and EDS analysis of $Cu_xTiS_2$ with different concentrations

Figure S5: (a) XRD pattern of pristine $NbS_2$ and saturated $Cu_{0.5}NbS_2$; (b-c) SEM image and elemental mapping of $Cu_{0.5}NbS$

Figure S6: (a) XRD pattern of pristine $TaS_2$ and saturated $Cu_{0.5}TaS_2$; (b-c) SEM image and elemental mapping of $Cu_{0.5}TaS_2$

Figure S7: (a) XRD pattern of pristine $TiSe_2$ and saturated $Cu_{0.5}TiSe_2$; (b-c) SEM image and elemental mapping of $Cu_{0.5}TiSe_2$

Figure S8: (a) XRD pattern of pristine $NbSe_2$ and saturated $Cu_{0.5}NbSe_2$; (b-c) SEM image and elemental mapping of $Cu_{0.5}NbSe_2$

Figure S9: (a) XRD pattern of pristine $TaSe_2$ and saturated $Cu_{0.5}TaSe_2$; (b-c) SEM image and elemental mapping of $Cu_{0.2}TaSe_2$

Figure S10: (a-b) SEM image and elemental mapping of saturated Mn intercalated $TiS_2$; (c) EDS analysis of saturated Mn intercalated $TiS_2$, which indicate the saturated concentration of Mn is $Mn_{0.66}TiS_2$

Figure S11: (a-b) SEM image and elemental mapping of saturated Fe intercalated $TiS_2$; (c) EDS analysis of saturated Fe intercalated $TiS_2$, which indicate the saturated concentration of Fe is $Fe_{0.66}TiS_2$

Figure S12: (a-b) SEM image and elemental mapping of saturated Co intercalated $TiS_2$; (c) EDS analysis of saturated Co intercalated $TiS_2$, which indicate the saturated concentration of Co is $Co_{0.66}TiS_2$

Figure S13: (a-b) SEM image and elemental mapping of saturated Ni intercalated $TiS_2$; (c) EDS analysis of saturated Ni intercalated $TiS_2$, which indicate the saturated concentration of Ni is $Ni_{0.66}TiS_2$

Figure S14: (a-b) SEM image and elemental mapping of saturated Cu intercalated $TiS_2$; (c) EDS analysis of saturated Cu intercalated $TiS_2$, which indicate the saturated concentration of Cu is $Cu_{0.5}TiS_2$

Figure S15: (a-b) SEM image and elemental mapping of saturated Ag intercalated $TiS_2$; (c) EDS analysis of saturated Ag intercalated $TiS_2$, which indicate the saturated concentration of Ag is $Ag_{0.5}TiS_2$

Figure S16: (a) XRD pattern of pristine $Ti_{0.5}V_{0.5}S_2$ and $Cu_xTi_{0.5}V_{0.5}S_2$ ($x$=0.05, 0.125, 0.25, 0.5); (b-c) EDS analysis and SEM image of $Cu_xTiS_2$ with different concentrations

Figure S17: Calculated project density of states (PDOS) of Ti in (a) $TiS_2$ and (b) $Ti_{0.5}V_{0.5}S_2$

Figure S18: Side views of charge density difference of $Cu_{0.25}TiS_2$ with $[01\bar{1}0]$ viewing direction, yellow region: charge accumulation; blue region: charge depletion

## Acknowledgments

This study was financially supported by the National Natural Science Foundation of China (Grant No. 52172254 and U2004212), Ningbo Natural Science Foundation (Grant No. 2022J297), International Partnership Program of Chinese Academy of Sciences (Grant No. 174433KYSB20190019), Leading Innovative and Entrepreneur Team Introduction Program of Zhejiang (Grant No. 2019R01003), Key R & D Projects of Zhejiang Province (Grant No. 2022C01236), Doctoral Training Partnership (DTP) from University of Nottingham Ningbo China. Mian Li also acknowledge the supporting of Youth Innovation Promotion Association CAS (Grant No. 2022298). We also thank Xiaofei Hu (Center of Test and Analysis, Ningbo Institute of Materials

Technology and Engineering, Chinese Academy of Sciences, China ) for TEM data and simulation.

## Author contributions

Qing Huang and Mian Li conceived the central idea of this study. Lin Gao and Mian Li designed the synthesis of experiments. Lin Gao performed all the experiment and characterization. Binjie Hu and Lin Gao helped the VASP calculation for the host modulation part. Lin Gao analyzed the data and wrote this paper. Qing Huang, Mian Li and Binjie Hu supervised the research and carefully modified the manuscript. All authors participated in the discussion of the data analysis and commented on the paper.

## Declaration of interests

The authors declare no competing interests.